\documentclass{PoS}

\title{Gluon Propagators in Linear Covariant Gauge}

\ShortTitle{Linear Covariant Gauge}

\author{\speaker{Attilio Cucchieri}$^{ab}$, Tereza Mendes$^b$,
        Gilberto M.\ Nakamura$^b$ and Elton M.\ S.\ Santos$^{bc}$\\
\llap{$^a$}Ghent University, Department of Physics and Astronomy, \\
           Krijgslaan 281-S9, 9000 Gent, Belgium \\
\llap{$^b$}Instituto de F\'\i sica de S\~ao Carlos, Universidade de S\~ao Paulo, \\
           Caixa Postal 369, 13560-970 S\~ao Carlos, SP, Brazil \\
\llap{$^c$}Instituto de Educa\c c\~ao, Agricultura e Ambiente, Campus Vale do Rio Madeira, \\
           Universidade Federal do Amazonas, 69800-000 Humait\'a, AM, Brazil \\
E-mail: \email{attilio@ifsc.usp.br}, \email{mendes@ifsc.usp.br}, \email{gilberto@ifsc.usp.br},
        \email{elton@ifsc.usp.br}}

\abstract{
The implementation of the linear covariant gauge on the lattice faces a
conceptual problem: using the standard compact discretization, the gluon
field is bounded, while the four-divergence of the gluon field satisfies a
Gaussian distribution, i.e.\ it is unbounded. This can give rise to
convergence problems when a numerical implementation is attempted. In
order to overcome this problem, one can use different discretizations for
the gluon field or consider an SU($N_c$) group with sufficiently large $N_c$. 
One can also consider small values of the gauge parameter $\xi$
and study numerically the limiting case of $\xi \to 0$, i.e.\ the Landau
gauge. These different approaches will be discussed here.}

\FullConference{The many faces of QCD\\
		November 1-5, 2010\\
		Gent Belgium}

\usepackage{epsfig}

\newcommand{\ee}{\end{equation}} 
\newcommand{\be}{\begin{equation}} 
\newcommand{\ec}{\end{center}} 
\newcommand{\bc}{\begin{center}} 
\newcommand{\eea}{\end{eqnarray}} 
\newcommand{\bea}{\begin{eqnarray}} 
\newcommand{\bd}{\begin{description}} 
\newcommand{\ed}{\end{description}} 
\newcommand{\bi}{\begin{itemize}} 
\newcommand{\ei}{\end{itemize}}

\def\spose#1{\hbox to 0pt{#1\hss}}
\def\ltapprox{\mathrel{\spose{\lower 3pt\hbox{$\mathchar"218$}}
 \raise 2.0pt\hbox{$\mathchar"13C$}}}
\def\gtapprox{\mathrel{\spose{\lower 3pt\hbox{$\mathchar"218$}}
 \raise 2.0pt\hbox{$\mathchar"13E$}}}

\begin{document}

%%%%%%%%%%%%%%%%%%%%%%%%%%%%%%%%%%%%%%%%%%%%%%%%%%%%%%%%%%%%%%%%%%%%%%%%%%%%%%%%%%%%%%%%%%%%%%%%%%%%%%%%%

\section{Why study the linear covariant gauge?}

Extensive numerical simulations on very large lattices \cite{Bogolubsky:2007ud,
Cucchieri:2007md,Sternbeck:2007ug,Cucchieri:2007rg,Cucchieri:2008fc,
Bogolubsky:2009dc,Gong:2008td} have shown that, in three and in four space-time
dimensions, the Landau gluon propagator displays a massive solution at small momenta
and the Landau ghost propagator is essentially free in the same limit. (For a recent
review and various comments, see respectively \cite{Cucchieri:2010xr} and 
\cite{Cucchieri:2011um}.) These
results are not in agreement with the original Gribov-Zwanziger confinement scenario
\cite{Gribov:1977wm, Zwanziger:1993dh} but they can be explained in the so-called
{\em refined Gribov-Zwanziger framework} \cite{Dudal:2008sp,Sorella:2011tu}.

A natural extension of these works in Landau gauge would be to consider the
linear covariant gauge, which has the Landau gauge as a limiting case.
However, until recently, the numerical gauge fixing for the linear covariant gauge
\cite{Giusti:1996kf,Giusti:2000yc,Cucchieri:2008zx} was not satisfactory. In Reference
\cite{Cucchieri:2009kk} we introduced a new implementation of the linear covariant
gauge on the lattice that solves most problems encountered in earlier implementations.
The final goal is to evaluate Green's functions numerically in the 
linear covariant gauge and, in particular, in the Feynman gauge. 
This could allow a nonperturbative evaluation of the gauge-invariant off-shell
Green's functions of the pinch technique \cite{Binosi:2009qm,Cornwall:2009as}.

%%%%%%%%%%%%%%%%%%%%%%%%%%%%%%%%%%%%%%%%%%%%%%%%%%%%%%%%%%%%%%%%%%%%%%%%%%%%%%%%%%%%%%%%%%%%%%%%%%%%%%%%%

\section{The Linear Covariant Gauge on the Lattice}

In the continuum, the linear covariant gauge is obtained by imposing the gauge condition
\begin{equation}
\partial_{\mu} A^b_\mu(x)=\Lambda^b(x) \; ,
\label{eq:lincov}
\end{equation}
where the real-valued functions $\Lambda^b(x)$ are generated using a Gaussian distribution
\begin{equation}
     \exp{ \left\{ \, -\frac{1}{2\xi} \int d^dx \sum_b \left[\Lambda^b(x)\right]^2 \, \right\} }
\label{eq:gauss}
\end{equation}
with width $\sqrt{\xi}$. The limiting case $\xi \to 0$, which implies $\Lambda^b(x)=0$, yields
the Landau gauge.

On the lattice, one can fix the Landau gauge by minimizing the functional
\begin{equation}
{\cal E}_{LG}[U^{g}] = - \mbox{Tr} \sum_{\mu, x} g(x) U_{\mu}(x) g^{\dagger}(x+e_{\mu})
\label{eq:ELG}
\end{equation}
with respect to the gauge transformations $\{g(x)\}$. Here, $U_{\mu}(x)$ are (fixed) link
variables and $g(x)$ are site variables, both belonging to the SU($N_c$) group. The sum is
taken over all lattice sites $x$ and directions $\mu$, and $\mbox{Tr}$ indicates trace in
color space. By considering a one-parameter subgroup\footnote{Here we indicate with
$\lambda^{b}$ a basis for the SU($N_c$) Lie algebra and with $\gamma^{b}(x)$ any
real-valued function.}
\begin{equation}
g(x,\tau) \, = \, \exp \left[ i \tau \gamma^{b}(x) \lambda^{b} \right] 
\end{equation}
of the gauge transformation $\{ g(x) \}$, one can verify that the stationarity condition
for the functional ${\cal E}_{LG}[U^{g}]$ implies the (lattice) gauge condition
\begin{equation}
 \sum_{\mu} \, A_{\mu}^{b}(x) \, - \, A_{\mu}^{b}(x-e_{\mu}) \, = \, 0 \; ,
\label{eq:Landaulatt}
\end{equation}
where
\begin{equation}
A_{\mu}(x) \, = \, ( \, 2 i \,)^{-1} \, \left[ U_{\mu}(x) - U_{\mu}^{\dagger}(x) \right]_{traceless}
\label{eq:discret}
\end{equation}
is the usual lattice discretization for the gluon field. Also, from the second variation of 
${\cal E}_{LG}[U^{g}]$ one can obtain a discretized version of the Faddeev-Popov operator
\begin{equation}
{\cal M}^{ab} \,=\, - D_{\mu}^{ab} \, \partial_{\mu} \; ,
\end{equation}
where $\, D_{\mu}^{ab} \,$ is the covariant derivative. Clearly, for the gauge-fixed
configurations, i.e.\ for all local minima of the functional ${\cal E}_{LG}[U^{g}]$,
this operator is positive-definite. This set of local minima defines the first Gribov
region $\Omega$ \cite{Gribov:1977wm,Zwanziger:1993dh}.

In Reference \cite{Cucchieri:2009kk} we have introduced the minimizing functional\footnote{It
is interesting to note that this functional can be interpreted as a spin-glass Hamiltonian
for the spin variables $g(x)$ with a random interaction given by $U_{\mu}(x)$ and with a
random external magnetic field $ \Lambda(x) $.}
\begin{equation}
{\cal E}_{LCG}[U^{g}, g, \Lambda] \; = \; {\cal E}_{LG}[U^{g}]
                   \, + \, \Re \; Tr \sum_x \,  i\, g(x) \, \Lambda(x) \; ,
\label{eq:ELCG}
\end{equation}
which is a natural extension of the Landau functional (\ref{eq:ELG}). Here $ \Re $ indicates
real part. One should stress that, in the numerical minimization, the link variables $U_{\mu}(x)$
are gauge-transformed to $g(x) U_{\mu}(x) g^{\dagger}(x+e_{\mu})$, while the $\Lambda^b(x)$
functions do not get modified. It is easy to verify, using again a one-parameter subgroup
$ g(x,\tau) $, that this functional leads to the lattice linear covariant gauge condition
\begin{equation}
\sum_{\mu} \, A^b_\mu(x) \,-\, A^b_\mu(x-e_{\mu}) \, = \, \Lambda^b(x) \; .
\label{eq:lincovlatt}
\end{equation}
Note that the above relation and periodic boundary conditions yield
\begin{equation}
\sum_x \Lambda^{b}(x) \, = \, 0 \; .
\end{equation}
This equality must be enforced numerically, within machine precision, when the functions
$ \Lambda^{b}(x)$ are generated using the Gaussian distribution (\ref{eq:gauss}).
Also note that the second variation (with respect to the parameter $\tau$) of the term
$\,i\, g(x) \, \Lambda(x)\,$ is purely imaginary and it does not contribute to the
Faddeev-Popov matrix ${\cal M}$. Clearly, having a minimizing functional for the
linear covariant gauge implies that the set of its local minima defines the first Gribov region
$\Omega$ and that the corresponding Faddeev-Popov operator ${\cal M}$ is positive-definite.
This should allow the extension of the Gribov-Zwanziger approach to the linear covariant gauge.
In particular, one should be able to study the region $\Omega$ for the case of a gauge parameter
$\xi \neq 0$ and to compare the results with the analytic study carried out in Reference
\cite{Sobreiro:2005vn} for small values of $\xi$.

In order to relate the lattice approach to the continuum, one should note \cite{Cucchieri:2008zx}
that the continuum relation (\ref{eq:lincov}) can be made dimensionless by
multiplying both sides by $a^2 g_0$. Since in $d$ dimensions and in the SU($N_c$) case one
has $\beta = 2 N_c / (a^{4-d} g_0^2)$, it is clear that the lattice quantity
\begin{equation}
\frac{\beta / (2 N_c)}{2\xi} \sum_{x, b} \left[ a^2 g_0 \Lambda^b(x)\right]^2
\, = \, \frac{1}{2\sigma^2} \sum_{x, b} \left[ a^2 g_0 \Lambda^b(x)\right]^2
\, = \, \frac{1}{2\sigma^2} \sum_{x, b} \left[ \Lambda_{latt}^b(x)\right]^2
\end{equation}
gives
\begin{equation}
 \frac{1}{2\xi} \frac{1}{a^{4-d} g_0^2} \int \frac{d^dx}{a^d}
     \sum_b \left[a^2 g_0 \Lambda^b(x)\right]^2 \;  = \;
     \frac{1}{2\xi} \int d^dx \sum_b \left[\Lambda^b(x)\right]^2 
\end{equation}
in the formal continuum limit. Thus, the continuum gauge parameter $\xi$ corresponds
to a width
\begin{equation}
\sigma \, = \, \sqrt{\frac{2 N_c \xi}{\beta} }
\label{eq:sigma}
\end{equation}
for the Gaussian distribution on the lattice and only for $\beta = 2 N_c$ does one have
$\sigma=\sqrt{\xi}$.

%%%%%%%%%%%%%%%%%%%%%%%%%%%%%%%%%%%%%%%%%%%%%%%%%%%%%%%%%%%%%%%%%%%%%%%%%%%%%%%%%%%%%%%%%%%%%%%%%%%%%%%%%

\section{Numerical Simulations}

The functional ${\cal E}_{LCG}[U^{g}, g, \Lambda]$ --- see Eqs.\ (\ref{eq:ELG}) and (\ref{eq:ELCG})
--- is linear in the gauge transformations $\{ g(x) \}$. Thus, the gauge-fixing algorithms usually
employed in the Landau case \cite{Cucchieri:1995pn,Cucchieri:1996jm,Cucchieri:2003fb} can be used
also for the linear covariant gauge and, in principle, the numerical gauge fixing should not be
a problem for this gauge. Nevertheless, as explained in the Abstract, any formulation
of the linear covariant gauge on the lattice faces a conceptual problem. Indeed, the gluon field
$A^a_\mu(x)$, usually defined as in Equation (\ref{eq:discret}), is bounded while the functions $\Lambda^b(x)$
satisfy a Gaussian distribution [see Equation (\ref{eq:gauss})] and are thus unbounded. Then, it is
clear that Eq.\ (\ref{eq:lincovlatt}) cannot be satisfied if $\Lambda^b(x)$ is too large \cite{Rank}.
As a consequence, one has to deal with convergence problems when numerically imposing the linear
covariant gauge condition. Also, these problems become more severe as the width $\sqrt{\xi}$ of the
Gaussian distribution becomes larger and/or as the lattice volume becomes larger. On the lattice, as shown
above, the width $\sigma$ of the Gaussian distribution is given by $\sigma=\sqrt{2 N_c \xi/\beta}$.
Thus, these convergence problems are more severe also for small values of the coupling $\beta$.
In particular, for $\beta < 2 N_c$ the lattice width $\sigma$ is larger than the continuum width
$\sqrt{\xi}$.

In References \cite{Cucchieri:2009kk,Cucchieri:2010ku} we have presented tests of convergence
of the numerical gauge fixing in the SU(2) case for relatively small lattice volumes, $\beta = 4$
and values of $\xi$ up to 0.5. [Recall that for $\beta = 4$ one has $\sigma=\sqrt{\xi}$ in the
SU(2) case.] We have also checked that the quantity $D_L(p^2) p^2$, where $D_L(p^2)$ is the
longitudinal gluon propagator, is approximately constant for all cases considered, as predicted
by Slavnov-Taylor identities. This verification failed in previous formulations of the lattice
linear covariant gauge \cite{Giusti:2000yc,Cucchieri:2008zx}.

In order to overcome the convergence problems discussed above and be able to simulate at lattice
coupling $\beta$ smaller than $4$ in the SU(2) case, we considered different discretizations of
the gluon field. In particular, we used the angle (or logarithmic) projection \cite{Amemiya:1998jz}
and the stereographic projection \cite{vonSmekal:2007ns} (for a slightly different implementation
of the stereographic projection see also \cite{Gutbrod:2004qp}). Note that, in the latter case,
the gluon field is unbounded even for a finite lattice spacing $a$. Our results \cite{Cucchieri:2009kk,
Cucchieri:2010ku} clearly show that the angle projection is already an improvement compared to the
standard discretization and that the best convergence is obtained when using the stereographic projection.
In References \cite{Cucchieri:2009kk,Cucchieri:2010ku} we also presented preliminary results for the
transverse gluon propagator using the stereographic projection. From these results one clearly sees
that, as in Landau gauge, the transverse propagator is more infrared suppressed when the lattice
volume increases. At the same time, for a fixed volume $V$, this propagator is also more infrared
suppressed when the gauge parameter $\xi$ increases. The latter result is in agreement with Reference
\cite{Giusti:2000yc}. One should, however, stress that the stereographic projection cannot
be extended to SU($N_c$) groups with $N_c > 2$. Thus, in the SU(3) case one should probably rely
on the logarithmic projection.

Finally, one should note that, in the SU(2) case, the value $\sigma = \sqrt{\xi}$, i.e.\ $\beta = 4$,
corresponds to a lattice spacing $a \approx 0.001$ fm. On the contrary, in the SU(3) case, one has
$\sigma = \sqrt{\xi}$ for $\beta = 6$, corresponding to $a = 0.102$ fm. Also, for a fixed t'Hooft
coupling $g_0^2 N_c$, we have $\beta \propto N_c^2$ and $\sigma \propto \sqrt{1/N_c}$. This suggests
that simulations for the linear covariant gauge are easier in the SU($N_c$) case for large $N_c$.
In Reference \cite{Cucchieri:2011aa} we tested this hypothesis by simulating the SU(2), SU(3) and SU(4)
cases for a gauge parameter $\xi=1$ and lattice volumes up to $V=32^4$ for several values of the
coupling $\beta$. We find that the convergence problems are indeed reduced when the number of colors
$N_c$ is larger.

%%%%%%%%%%%%%%%%%%%%%%%%%%%%%%%%%%%%%%%%%%%%%%%%%%%%%%%%%%%%%%%%%%%%%%%%%%%%%%%%%%%%%%%%%%%%%%%%%%%%%%%%%

\section{The Limit $\xi \to 0$}

In the continuum, the Landau gauge condition is defined by considering the usual Faddeev-Popov
Lagrangian for the linear covariant gauge and by taking the limit $\xi \to 0$. On the contrary,
on the lattice, this gauge has been studied without considering this limit, but by imposing
directly the gauge condition
\begin{equation}
\partial_{\mu} \, A_{\mu}^b(x) \, = \, 0 \, ,
\label{eq:lorenz}
\end{equation}
which is usually called the Lorenz gauge. The latter gauge condition is fixed numerically by
minimizing the functional (\ref{eq:ELG}). Using our implementation of the linear covariant gauge, it
seems natural to study the Landau gauge by numerically considering the limit $\xi \to 0$, in analogy
with the definition in the continuum. One should also note that, in this limit, the width of the
Gaussian distribution (\ref{eq:gauss}) goes to zero and, therefore, the convergence problems discussed
above should be reduced --- or eliminated --- even for large lattice volumes and for $\beta$ values in
the scaling region. Moreover, since in the limit $\xi \to 0$ the Gaussian distribution becomes a
Dirac delta function $\delta(\Lambda)$, this limit can be studied numerically by using
different approximations of the delta function. 

Here we consider three possible sequences of functions, labelled by a parameter $\alpha$, leading
to a delta function in the limit $\alpha \to 0$, i.e.
\begin{enumerate}
\item the Gaussian distribution 
\begin{equation}
f_G(\Lambda) \, = \, \frac{ e^{-\Lambda^2/ (2 \alpha_G^2)} }{ \sqrt{2\pi \alpha_G^2} } \, ,
\end{equation}
\item the Triangle distribution 
\begin{equation}
f_T(\Lambda) \, = \, \left\{ \begin{array}{l l}
              \frac{(1 - \Lambda/\alpha_T)}{\alpha_T} & \quad \mbox{for} \, \Lambda \in [0,\alpha_T] \\[2mm]
              \frac{(1 + \Lambda/\alpha_T)}{\alpha_T} & \quad \mbox{for} \, \Lambda \in [-\alpha_T,0] \\[2mm]
                                                    0 & \quad \mbox{for} \, \Lambda \notin [-\alpha_T,\alpha_T] \\
                             \end{array} \right.
\end{equation}
\item and the Rectangular distribution
\begin{equation}
f_R(\Lambda) \, = \, \left\{ \begin{array}{l l}
                 \frac{1}{\alpha_R} & \quad \mbox{for} \,\Lambda \in [-\alpha_R/2,\alpha_R/2] \\[2mm]
                                 0  & \quad \mbox{for} \,\Lambda \notin [-\alpha_R/2,\alpha_R/2] \\
                               \end{array} \right. \, .
\end{equation}
\end{enumerate}
Note that the last two distributions are bounded. Also note that the average value of
$ \Lambda^2 $ using these three distributions is respectively given by $\alpha_G^2$,
$\alpha_T^2/6$ and $\alpha_R^2/12$. Thus, by setting 
\begin{equation}
\alpha_G \, = \, \frac{\alpha_T}{\sqrt{6}} \, = \, \frac{\alpha_R}{\sqrt{12}}
\label{eq:alphas}
\end{equation}
we have three different distributions with the same moment of inertia.

\begin{figure}
\begin{center}
\vspace{-2cm}
\includegraphics[width=.70\textwidth]{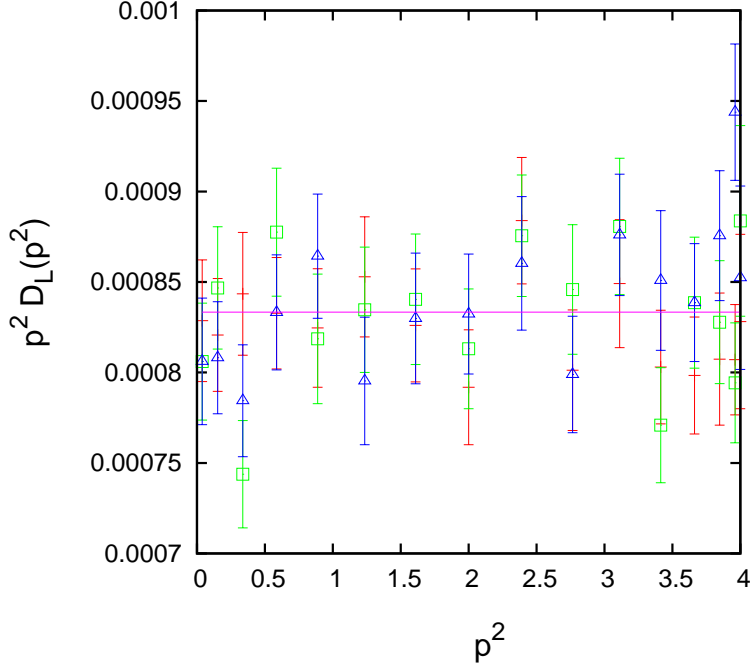} 
\vspace{-3cm}
\caption{Plot of $p^2 D_L(p^2)$ as a function of $p^2$ (in lattice units) for
$\alpha_R = 0.1$ ($+$),
$\alpha_T \approx 0.0707107$ ($\square$) and
$\alpha_G \approx 0.0288674$ ($\triangle$).
We also show the theoretical value $\sigma^2 = \alpha_R^2 / 12 \approx 0.0008333$.}
\label{fig:lon-0.1}
\end{center}
\end{figure}

\begin{figure}
\begin{center}
\vspace{-3cm}
\includegraphics[width=.70\textwidth]{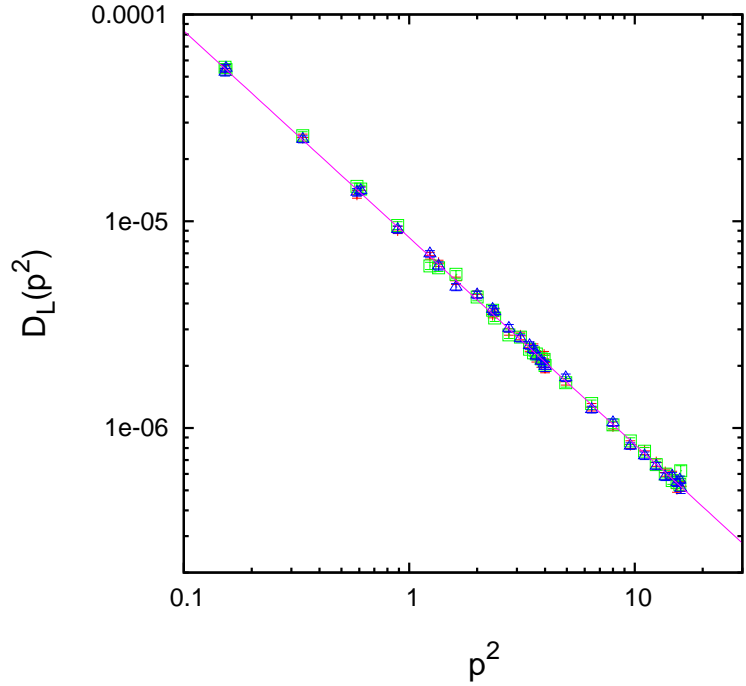} 
\vspace{-3cm}
\caption{Plot of $D_L(p^2)$ as a function of $p^2$ (both in lattice units), using the logarithmic scale
in both axes, for
$\alpha_R = 0.01$ ($+$),
$\alpha_T \approx 0.00707107$ ($\square$) and
$\alpha_G \approx 0.00288674$ ($\triangle$).
We also show the theoretical prediction $\alpha_R^2 / (12 p^2)$.}
\vspace{-1cm}
\label{fig:lon-0.01}
\end{center}
\end{figure}

\begin{figure}
\begin{center}
\vspace{-3cm}
\includegraphics[width=.65\textwidth]{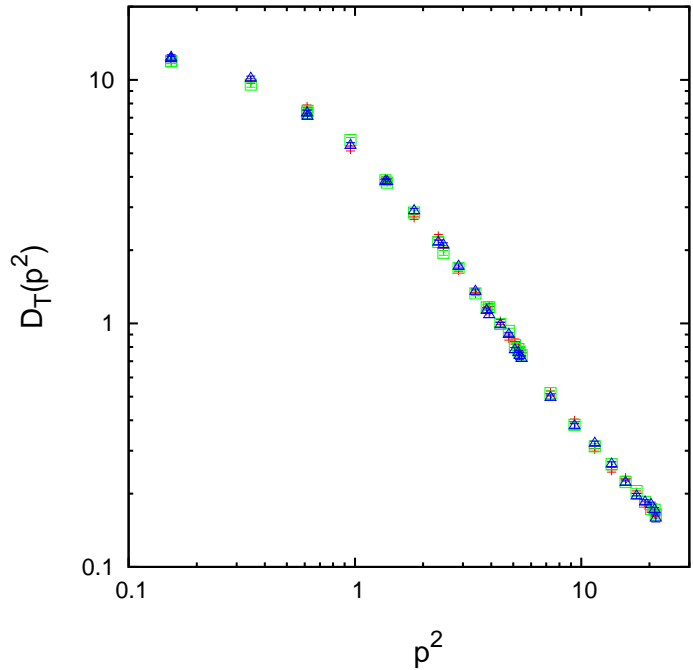}
\vspace{-2.5cm}
\caption{Plot of $D_T(p^2)$ as a function of $p^2$ (both in lattice units), using the logarithmic scale
in both axes, for
$\alpha_R = 0.1$ ($+$),
$\alpha_T \approx 0.0707107$ ($\square$) and
$\alpha_G \approx 0.0288674$ ($\triangle$).}
\vspace{-1cm}
\label{fig:tra-0.1}
\end{center}
\end{figure}

\begin{figure}
\begin{center}
\vspace{-2cm}
\includegraphics[width=.67\textwidth]{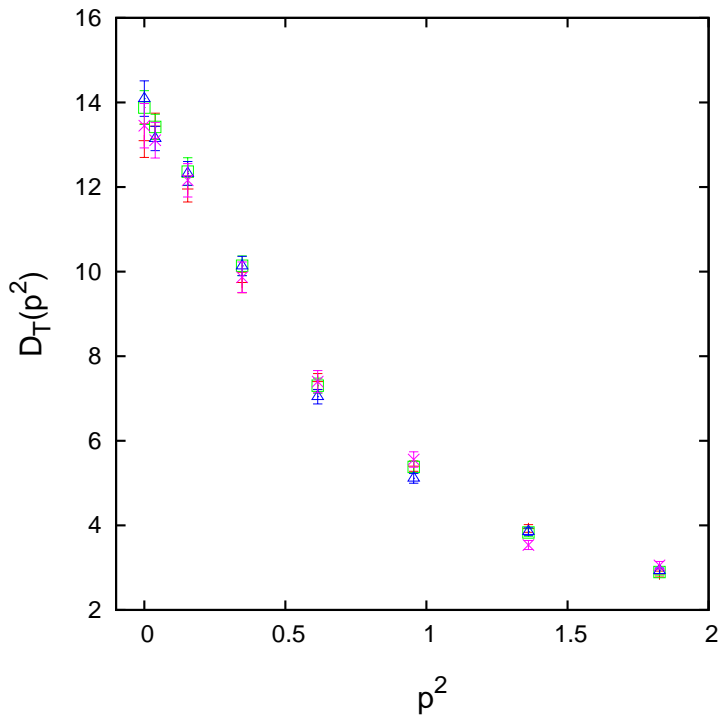}
\vspace{-2.5cm}
\caption{Plot of $D_T(p^2)$ as a function of $p^2$ (both in lattice units) for
$\alpha_G \approx 0.144338$ ($+$),
$\alpha_G \approx 0.0288674$ ($\square$),
$\alpha_G \approx 0.00288674$ ($\triangle$) and
$\alpha_G = 0$ ($\times$).}
\label{fig:tra}
\end{center}
\end{figure}

We have done exploratory tests considering, in the SU(2) case, the lattice volume $V = 32^4$ at
$\beta = 2.2$. For the gauge-fixing parameters $\alpha$ we used the values $\alpha_G \approx 0.144338,
0.0288674$ and $0.00288674$ for the Gaussian distribution, $\alpha_T \approx 0.353553, 0.0707107$ and $0.00707107$
for the triangle distribution and $\alpha_R = 0.5, 0.1$ and $0.01$ for the rectangular distribution.
Note that these values satisfy the relations in Eq.\ (\ref{eq:alphas}). Also, the values
of $\alpha_G$ correspond to the continuum values $\xi \approx 0.0114584, 4.5833 \, 10^{-4}$ and $4.5833 \,
10^{-6}$, respectively. For comparison, we have also done simulations directly at $\xi = 0$, i.e.\
imposing the Lorenz condition (\ref{eq:lorenz}). Results for the longitudinal $D_L(p^2)$ and the
transverse $D_T(p^2)$ gluon propagators are shown in Figs.\ \ref{fig:lon-0.1}--\ref{fig:tra}.
One sees that the limit $\alpha \to 0$ is smoothly approached in the gluon sector and that
the results are independent of the considered distribution. In particular, it is clear in Figs.\
\ref{fig:lon-0.1}--\ref{fig:lon-0.01} that the theoretical prediction $D_L(p^2) = \sigma^2 / p^2$,
with $\sigma^2 = \alpha_G^2 = \alpha_R^2 / 12$, is satisfied by the data obtained with these three
distributions. At the same, a value of $\sigma^2 = \alpha_G^2 \approx 0.144338^2 \approx 0.02$ (see Fig.\
\ref{fig:tra}), which corresponds to the continuum value $\xi \approx 0.01$,
already seems to give results in quantitative agreement with the Landau case.

%%%%%%%%%%%%%%%%%%%%%%%%%%%%%%%%%%%%%%%%%%%%%%%%%%%%%%%%%%%%%%%%%%%%%%%%%%%%%%%%%%%%%%%%%%%%%%%%%%%%%%%%%

\section{Conclusions}

We have found a minimizing functional for the linear covariant gauge that is a simple generalization
of the Landau-gauge functional. This approach solves most problems encountered in earlier
implementations. Simulations for large lattice volumes, $\beta$ values in the scaling region and a
large gauge parameter $\xi$ can probably be done with SU(3) and SU(4) using the logarithmic projection,
allowing a non-perturbative study of Green's functions. Finally, the approach to the limiting case
$\xi \to 0$, i.e. the Landau gauge, can also be studied numerically. Here we have investigated this
limit considering three different distributions defining the gauge condition.

%%%%%%%%%%%%%%%%%%%%%%%%%%%%%%%%%%%%%%%%%%%%%%%%%%%%%%%%%%%%%%%%%%%%%%%%%%%%%%%%%%%%%%%%%%%%%%%%%%%%%%%%%

\section*{Acknowledgments}
We thank Matthieu Tissier for helpful discussions and the organizers of {\em The Many Faces of QCD}
for a very pleasant and stimulating workshop.
This work has been partially supported by the Brazilian agencies FAPESP, CNPq and
CAPES. In particular, support from FAPESP (under grant \# 2009/50180-0) is acknowledged.

%%%%%%%%%%%%%%%%%%%%%%%%%%%%%%%%%%%%%%%%%%%%%%%%%%%%%%%%%%%%%%%%%%%%%%%%%%%%%%%%%%%%%%%%%%%%%%%%%%%%%%%%%

\end{document}